\def\spose#1{\hbox to 0pt{#1\hss}}

\def\multleft#1{\hbox to size{\vbox {\halign {\lft{##}\cr #1}}\hfill}\par}
\def\multright#1{\hbox to size{\vbox {\halign {\rt{##}\cr #1}}\hfill}\par}

\def\today{\ifcase\month\or January\or February\or March\or April\or May\or
      June\or July\or August\or September\or October\or November\or December\fi
      \space\number\day, \number\year}
\def\s{\hbox{\phantom{5}}}	

\def\km{{\rm\thinspace km}}

\def\Mpc{{\rm\thinspace Mpc}}

\def\s{{\rm\thinspace s}}

\def\kmps{\hbox{$\km\s^{-1}\,$}}

\def\kmpspMpc{\hbox{$\kmps\Mpc^{-1}$}}

\def\H2{\hbox{H$_{2}$}}

\newcommand{\gtsim}{\mbox{{\raisebox{-0.4ex}{$\stackrel{>}{{\scriptstyle\sim}}
$}}}}
\newcommand{\ltsim}{\mbox{{\raisebox{-0.4ex}{$\stackrel{<}{{\scriptstyle\sim}}
$}}}}
\newcommand{\mc}{\multicolumn}
\documentclass[usegraphicx]{mn2e}
\usepackage{latexsym,subfigure,epsfig,graphics}

\include{defn}

\begin{document}
\hsize=6truein

\title[Orientation dependency of broad-line widths in
quasars]{Orientation dependency of broad-line widths
in quasars and consequences for black-hole mass estimation}
\author[M.J.~Jarvis \& R.J.~McLure]
{Matt J.~Jarvis$^{1}$\thanks{Email: m.jarvis1@physics.ox.ac.uk} and Ross J.~McLure$^{2}$\thanks{Email: rjm@roe.ac.uk} \\
\footnotesize
$^{1}$Astrophysics, Department of Physics, Keble Road, Oxford, OX 3RH, U.K. \\
$^{2}$Institute for Astronomy, University of Edinburgh, Royal Observatory, Edinburgh EH9 3HJ, U.K.\\}

\maketitle

\begin{abstract}
In this paper we report new evidence that measurements of the
broad-line widths in quasars are dependent on the source orientation,
consistent with the idea that the broad-line region is flattened or
disc-like. This reinforces the view derived from radio-selected
samples, where the radio-core dominance has been used as a measure of
orientation. The results presented here show a highly significant ($>
99.95$ per cent) correlation between radio spectral index (which we
use as a proxy for source orientation) and broad-line width derived
from the H$\beta$ and MgII emission lines.  This is the first time
that this type of study has used quasars derived from a large
optically selected quasar sample, where the radio-loud quasars (RLQs)
and radio-quiet quasars (RQQs) have indistinguishable distributions in
redshift, bolometric luminosity and colour, and therefore overcomes
any biases which may be present in only selecting via radio emission.

We find that the mean FWHM for the flat-spectrum
($\alpha_{\rm rad} \ltsim 0.5$) radio-loud quasars (FSQs) to be $\overline{\rm FWHM}_{\rm FSQ} = 4990 \pm 536$~km~s$^{-1}$, which differs
significant from the mean FWHM of the steep-spectrum
($\alpha_{\rm rad} > 0.5$) radio-loud quasars (SSQs), where $\overline{{\rm FWHM}}_{\rm SSQ} = 6464 \pm 506$~km~s$^{-1}$. 
We also find that the distribution in FWHM for the FSQs is
indistinguishable from that of the radio-quiet quasars (RQQs), where $\overline{\rm FWHM}_{\rm RQQ} = 4831 \pm 25$~km~s$^{-1}$. Considering other observational results in the literature we interpret
this result in the context of a significant fraction of the FSQs being
derived from the underlying RQQ population which have their radio flux Doppler boosted above the RLQ/RQQ divide.

Under the assumption of a disc-like broad-line region we find no evidence for a difference in the average line-of-sight angle for RQQs and RLQs, implying that the difference is due to black-hole mass.
However, we caution
against the virial method to estimate black-hole masses in small or
ill-defined quasar samples due to significant orientation
dependencies.  Disentangling the relative importance of black-hole
mass and orientation would require higher resolution radio
observations. However, orientation effects could be
minimised by obtaining low frequency ($< 1$~GHz) radio observations of the optically selected SDSS quasars.

\end{abstract}

\section{INTRODUCTION}

The orientation dependent unification picture of AGN has now been in
place for well over a decade (Barthel 1989; Antonucci 1993) and in
particular, measures of source orientation in the radio waveband have
played a crucial role in the development of this picture (Urry \&
Padovani 1995). Under this unification scheme of radio-loud AGN, we
see the sources which have their radio jets pointing along our
line-of-sight as flat-radio-spectrum quasars (FSQs), with a radio spectral
index $\alpha_{\rm rad} < 0.5$\footnote{We use the definition $S_{\nu}
\propto \nu^{-\alpha_{\rm rad}}$, where $S_{\nu}$ is the flux density
at frequency $\nu$, and $\alpha_{\rm rad}$ is the radio spectral
index}. The flat spectrum arises from the superposition of many
relativistically beamed synchrotron self-absorbed spectra. These flat-spectrum sources usually exhibit
the optical characteristics of quasars, i.e. they are unresolved in
optical imaging observations and have broad permitted emission lines
(FWHM $> 2000$~km~s$^{-1}$). High-frequency ($\nu > 1$~GHz) radio
surveys (e.g. Drinkwater et al. 1997) preferentially pick-out
these FSQs due to the fact that their radio
emission does not become fainter at high frequencies, as is observed in
optically thin synchrotron sources.

Conversely, the other extreme is a source with its radio jets aligned
along the plane of the sky. In this case the bulk of the emission
arises from the optically thin lobes which can extend up to Mpc
scales, and these sources are classified as radio galaxies. The radio
spectra of the lobes which dominate the radio spectrum of these
sources are predominantly steep with $\alpha_{\rm rad} > 0.5$. These
sources dominate the low-frequency radio surveys, such as the
complete samples derived from the 3CRR (Laing, Riley \& Longair 1983), 6CE (Eales et al. 1997) 7CRS (Willott et al. 2002) and the Westerbork
Northern Sky Survey (WENSS; Rengelink et al. 1997). Optically these
sources appear as normal galaxies, usually with some extended line
emission in the direction of the radio jets; the so-called
alignment effect (e.g. van Breugel, Heckman \& Miley 1984; Inskip et al. 2002). The spectra of these objects only
contain narrow emission lines (FWHM $< 2000$~km~s$^{-1}$) that are
believed to reside beyond the obscuring torus.

Between these two extreme cases, where the jets lie at an angle
$\ltsim 45$~degrees (e.g. Barthel 1989) to the line-of-sight we
still see broad emission lines and also some emission from the flat-spectrum core, however there is
also a large contribution from the large scale, steep-spectrum radio
lobes. In this case the total radio spectra are somewhere in between
the flat-spectrum core dominated sources and the core-less,
lobe-dominated radio galaxies and can thus exhibit both components,
these are the steep-spectrum radio-loud quasars (SSQs).

Using this argument, the total radio spectral index is therefore an
indicator of the orientation of the source on the sky. This is also
reinforced by another orientation estimator based around the same
argument. The strength of the core radio emission in quasars, when
compared to the extended emission, has been used extensively as an indication
of source orientation (e.g. Wills \& Browne 1986; Brotherton 1996).

With these measures of orientation which are presented by radio-loud
sources we are able to probe the geometry of other properties inherent
to powerful quasars. One such property is the geometry of the
broad-line region.  Thus far, studies of this type have been
restricted to radio-selected samples which are difficult to reconcile
with the more abundant radio-quiet quasar population. However, with
the recent large-scale quasar surveys coupled with all-sky radio
surveys, we are now able to construct large samples of both radio-quiet
and radio-loud quasars with consistent selection criteria. 

In this paper we use a sub-sample of the Sloan Digital Sky Survey
quasar catalogue to investigate the dependency of broad-line width on
source orientation in a sample of radio-loud quasars. The important
difference of this study to previous studies is that the radio-loud
sources are selected with exactly the same criteria as the larger
population of radio-quiet sources in the sample. Thus we are able to
directly infer the orientation effects which may present a distorted
view of the radio-quiet population. Such an effect may be very
important if we are to continue using the broad-line measurements as
an important component in determining black-hole masses of quasars,
via the virial estimate (e.g. Kaspi et al. 2000; Vestergaard 2002;
McLure \& Jarvis 2002), over the history of the Universe (e.g. McLure \& Dunlop 2004).

The paper is set out as follows, in Section~\ref{sec:sample} we
provide a brief description of the sample of quasars defined by McLure
\& Jarvis (2004). In section~\ref{sec:orient} we present evidence for
a correlation between broad line width and radio spectral index, and
then go on to determine the parameters of a disc-like geometry for the broad-line region in RLQs.
In section~\ref{sec:discuss} we attempt to relate the broad-line
width distribution in the radio-loud quasars (RLQs) with that of the radio-quiet quasars (RQQs), and outline the implications that this would have
in using broad-line widths to estimate black-hole masses. We summarise
our results in section~\ref{sec:summary}.
All
cosmological calculations presented in this paper assume $\Omega_{\rm
M} = 0.3$, $\Omega_{\Lambda} = 0.7$, $H_{\circ} = 70$~\kmpspMpc.

\section{The sample}\label{sec:sample}

For this analysis we use quasars drawn from the full sample of McLure
\& Jarvis (2004). These were selected from the SDSS quasar catalogue
II (SQCII; Schneider et al. 2003). Only those quasars which fell
within the SDSS/FIRST overlap region were included\footnote{We note
that using the quasar catalogue from SDSS DR3 does not enhance our
data set in any way as there are the same number of radio-loud quasars
with our selection criteria.}. This ensures complete coverage in the
radio wave band and both
radio-loud and radio-quiet quasars were completely matched in terms of
optical luminosity and redshift distributions. Therefore, any biases
that could be present due to targeted selection of FIRST (Becker, White \& Helfand 1995) radio sources are negated.

For the purposes of this paper we use the definition of radio-loudness
of Miller, Peacock \& Mead (1990; i.e. $L_{1.4} >
10^{24}$~W~Hz$^{-1}$~sr$^{-1}$), which results in a total sample size
of 409. However, to ensure that we have an indicator of orientation
through measurement of the radio spectral index, we
cross-correlate our sample with the WENSS (Rengelink et
al. 1997), and to ensure a matching of spatial resolution also with
the NVSS (Condon et al. 1998). Both the NVSS and WENSS have spatial
resolutions of $\sim 30-40$~arcsec at 325~MHz and 1.4~GHz
respectively.

The need for each source to be matched with a source in the WENSS
means that the sample size is reduced significantly due to the fact
WENSS only covers the sky north of 30 degrees in declination and at a
significantly higher flux-density limit of $S_{325} > 18$~mJy
($5\sigma$ rms). Of the 409 objects in the radio-loud sample, 121 lie in
the area covered by the WENSS, 53 of which have flux-densities greater
the WENSS flux-density limit.

\section{Orientation dependence of broad-emission lines?}\label{sec:orient}

\subsection{Radio spectral index as a proxy for source orientation}\label{sec:proxy}

The radio sources in our sample are derived from the optically
selected SDSS quasar catalogue. Therefore in-depth studies have not
yet investigated the radio structure of these sources, unlike other
widely used radio surveys, such as the 3CRR (Laing et al. 1983).
For this reason we are limited in our analysis, due to the
fact that we have no morphological constraint on the source
orientation. Even the FIRST radio survey does
not have enough spatial resolution (the spatial resolution of FIRST is
$\sim 5$~arcsec) to measure the core-to-lobe flux ratio (the $R$
parameter; Hine \& Scheuer 1980; Orr \& Browne 1982; Hough \& Readhead 1989) for the
majority of our sample.

\begin{figure}
\includegraphics[width=0.48\textwidth,angle=0]{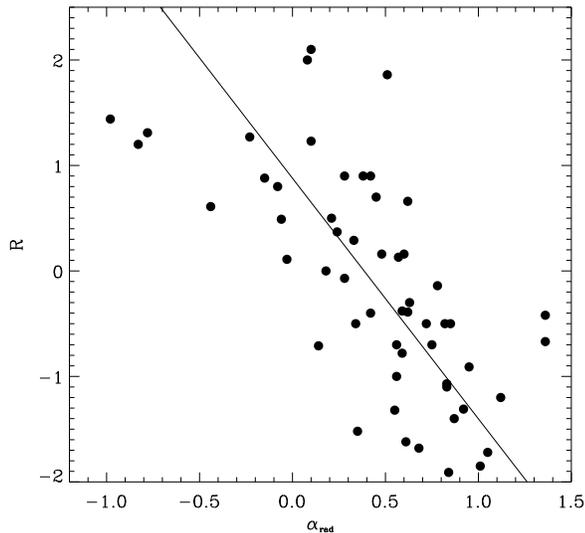}
\caption{The relation between the radio-core dominance
and the radio spectral index from the sample defined by Brotherton
(1996). The solid line is the best fit to the data where the fitting
takes account of uncertainties in both $\alpha_{\rm rad}$ and $R$.  }
\label{fig:broth}
\end{figure}

However, as previously discussed, many studies in the literature have shown that the
radio-spectral index $\alpha_{\rm rad}$ may be a good indicator of the
source orientation. This stems from the unification picture of
radio-loud AGN and has been confirmed observationally by both Wills \&
Browne (1986) and Brotherton (1996). Brotherton used a sample 60
radio-loud quasars to investigate the relation between radio-core
dominance and broad-line FWHM. In Fig.~\ref{fig:broth} we plot the
radio spectral index and the radio core-dominance parameter $R$, from
Table~1 of Brotherton (1996). This shows a highly significant
correlation ($>> 99.99$ per cent) between these two parameters with a
dispersion in $\alpha_{\rm rad}$ of $\sigma_{\alpha} = 0.37$ for a
given $R$ parameter (and $\sigma_{R} = 0.87$ for a given $\alpha_{\rm rad}$). Given that the $R$ parameter has been used
extensively to estimate the orientation of radio sources we use $\alpha_{\rm rad}$ as a proxy for $R$ and thus source orientation for the purposes of this study.

\begin{table}
\begin{center}
{\caption{\label{tab:sample} Radio-loud sources in the sample used
in this analysis. All radio luminosities are given in
W~Hz$^{-2}$~sr$^{-1}$ and the survey from which they are derived are
also listed.}}
\begin{tabular}{cccccc}
\mc{1}{c}{$z$} & \mc{1}{c}{$\alpha_{\rm rad}$} &
\mc{1}{c}{FWHM} & \mc{1}{c}{$\log_{10}(L_{325})$ } & \mc{1}{c}{$\log_{10}(L_{1.4})$ } & \mc{1}{c}{$\log_{10}(L_{1.4})$ } \\

\mc{1}{c}{} & \mc{1}{c}{} &
\mc{1}{c}{km~s$^{-1}$} & \mc{1}{c}{WENSS} & \mc{1}{c}{NVSS} & \mc{1}{c}{FIRST} \\
\hline\hline
 0.32&  0.53&     5333& 24.03& 23.51& 22.72\\
 0.34&  0.60&     3304& 25.31& 24.87& 24.63\\
 0.34&  0.90&    10531& 25.59& 24.75& 23.72\\
 0.41&  0.42&     2554& 24.49& 24.20& 24.12\\
 0.43&  0.64&     1902& 24.71& 24.29& 24.24\\
 0.45&  0.77&     9404& 25.01& 24.50& 24.46\\
 0.56&  0.51&     5695& 24.85& 24.44& 24.22\\
 0.58&  0.51&     5655& 25.68& 25.32& 25.22\\
 0.59&  0.23&     4836& 26.93& 26.76& 26.71\\
 0.64&  0.59&     2549& 24.62& 24.29& 24.37\\
 0.69& -0.04&     2478& 24.80& 24.85& 24.91\\
 0.70&  0.60&     6201& 25.20& 24.59& 24.20\\
 0.73& -0.92&     4082& 24.57& 25.02& 24.81\\
 0.77&  0.43&     5613& 25.17& 24.89& 24.87\\
 0.79&  1.26&     4831& 24.69& 24.31& 24.93\\
 0.81&  0.63&     6491& 25.77& 25.10& 24.71\\
 0.82&  0.80&     4672& 26.36& 25.44& 24.84\\
 0.84&  0.34&     2699& 25.88& 25.60& 25.51\\
 0.85&  0.47&     3545& 26.05& 25.69& 25.60\\
 0.88&  0.48&     7358& 25.88& 25.56& 25.55\\
 0.89&  0.85&     7519& 25.59& 24.96& 24.83\\
 0.93& -0.13&     5001& 25.43& 25.47& 25.42\\
 1.00&  0.71&     5946& 26.21& 25.38& 24.97\\
 1.02& -0.35&     2336& 24.90& 25.13& 25.13\\
 1.03&  0.65&     5854& 26.15& 25.22& 24.68\\
 1.07&  1.23&     8722& 27.66& 26.85& 26.83\\
 1.14&  0.49&     5122& 25.68& 25.36& 25.36\\
 1.17&  0.77&     4986& 25.84& 25.11& 24.90\\
 1.17&  0.23&     3618& 25.21& 25.03& 25.00\\
 1.32&  0.45&     7065& 27.15& 26.76& 26.68\\
 1.33&  0.73&     6496& 26.64& 25.97& 25.82\\
 1.35&  0.55&     4678& 25.61& 25.22& 25.20\\
 1.39&  0.43&     8801& 25.69& 25.35& 25.31\\
 1.41&  0.87&     7020& 27.05& 25.86& 25.43\\
 1.44&  0.97&    15694& 27.78& 26.20& 25.58\\
 1.44&  0.22&    13409& 26.02& 25.86& 25.84\\
 1.46&  0.41&     7165& 25.97& 25.63& 25.59\\
 1.48&  0.77&     6122& 27.20& 26.69& 26.68\\
 1.55& -0.18&     3097& 25.08& 25.20& 25.20\\
 1.57&  0.71&     7898& 26.73& 25.92& 25.73\\
 1.59&  1.06&     9156& 26.56& 25.44& 25.20\\
 1.63&  0.21&     3109& 25.69& 25.46& 25.41\\
 1.71&  0.18&     3366& 26.71& 26.52& 26.49\\
 1.71&  0.53&     5504& 25.84& 25.16& 25.00\\
 1.72&  0.30&     5371& 25.57& 25.33& 25.31\\
 1.74&  0.47&     2848& 25.89& 25.64& 25.66\\
 1.79&  0.39&     2897& 26.13& 25.86& 25.85\\
 1.80&  0.87&     4716& 28.45& 26.76& 26.28\\
 1.85&  0.47&     6462& 25.71& 25.36& 25.34\\
 1.85&  0.67&     8208& 26.25& 25.60& 25.51\\
 1.91&  0.94&     5891& 27.35& 26.48& 26.38\\
 1.99&  0.36&     5473& 26.72& 26.41& 26.38\\
 2.03&  0.08&     6435& 26.06& 26.02& 26.03\\
\hline\hline
\end{tabular}
\end{center}
\end{table}

\begin{figure}
\includegraphics[width=0.48\textwidth,angle=0]{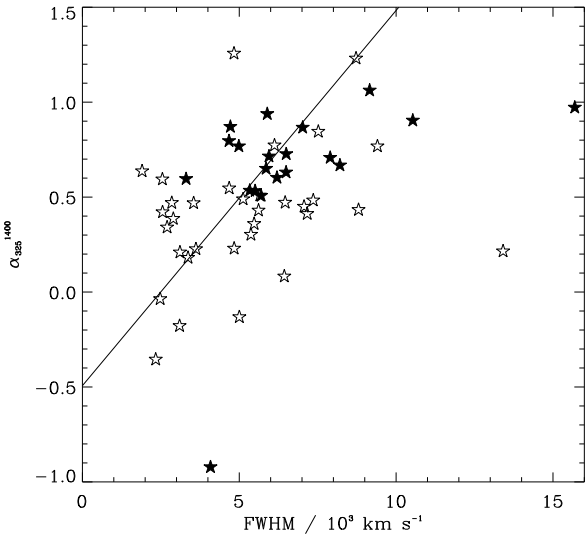}
\caption{The radio spectral index (measured between 1400 and 325~MHz)
versus the measured broad emission-line FWHM for the sample of
radio-loud quasars discussed in section~\ref{sec:sample}. The open
stars represent those sources where the integrated flux measured from
the FIRST catalogue is within a factor of two of the integrated flux
measured from the NVSS catalogue. The solid symbols are sources where
the ratio of FIRST/NVSS flux-density is less than two, implying that there is a significant extended component. The solid line represents the best fit to the correlation accounting for the uncertainties in both $\alpha_{\rm rad}$ and FWHM.
One can immediately see that the detection of extended emission is
heavily biased to those sources with steep ($\alpha_{\rm rad} > 0.5$) spectral index as expected in the orientation-based unification scheme. }
\label{fig:fig1}
\end{figure}

\subsection{FWHM versus radio spectral index}\label{sec:fwhmvsrad}

In Fig.~\ref{fig:fig1} we show the radio spectral index versus the
FWHM of the H$\beta$ or MgII emission lines, depending on
the redshift of the source ($0.1 < z < 0.7$ for H$\beta$ and $0.5 < z
< 2.1$ for MgII)\footnote{The method of fitting the
broad-line widths to these emission lines is described fully in McLure
\& Dunlop (2004), where they also show that the FWHM of H$\beta$ and
MgII are consistent for spectra where both are present.}.  The sample used here is comprised of all of those
sources with a detection in the WENSS catalogue (see Table~\ref{tab:sample}). One can immediately
see a correlation between the two parameters. The Spearman rank
coefficient for these two parameters is $\rho = 0.459$ with a
significance of the correlation being present $> 99.95$ per
cent. Using Kendall's rank we find a similar statistical significance
with $\tau=0.324$ and a significance again of $> 99.95$ per cent. As
these radio sources are selected by exactly the same method i.e. SDSS
optical selection and present in the WENSS, with no
biases that would influence the measurement of both the FWHM and
spectral index, this correlation is undoubtedly real.

Moreover, if we now split the radio-loud sources in this sample into
flat-spectrum ($\alpha_{\rm rad} < 0.5$; FSQs) and steep-spectrum
quasars ($\alpha_{\rm rad} > 0.5$; SSQs), then the difference in the
measured FWHM for these two populations is highly significantly. The
25 FSQs have a mean and standard error of FWHM $= 4990 \pm
536$~km~s$^{-1}$, whereas the 28 SSQs have a mean FWHM $= 6464 \pm
506$~km~s$^{-1}$. The KS-test also shows that the distributions in FWHM
for FSQs and SSQs are significantly different, with a probability of 4
per cent that they are drawn from the same underlying distribution.

This is in line with previous studies in the literature which have
used both the spectral index and the ratio of the core emission to the
extended lobe emission to investigate the dependence of broad-line
width on source orientation. Using a sample of radio-loud quasars
derived from various radio surveys Wills \& Browne (1986) found
evidence for a significant ($>99.9$ per cent) correlation between
radio-spectral index and the FWHM of broad emission lines. These
authors also compiled the necessary data to investigate the relation
between the $R$ parameter and the FWHM. Again they found a significant
relation between FWHM and $R$, and thus source orientation.  This can
be seen qualitatively in Fig.~\ref{fig:fig1}, where only the
steep-spectrum sources exhibit an excess of extended emission at the
NVSS resolution when compared to the higher resolution FIRST
flux-density.
The fact
that our sample is derived from an optically selected quasar sample
provides additional constraints on the link between orientation and
radio spectral index, and also a way of probing the distribution in orientation for RQQs.

\begin{figure}
\includegraphics[width=0.48\textwidth,angle=0]{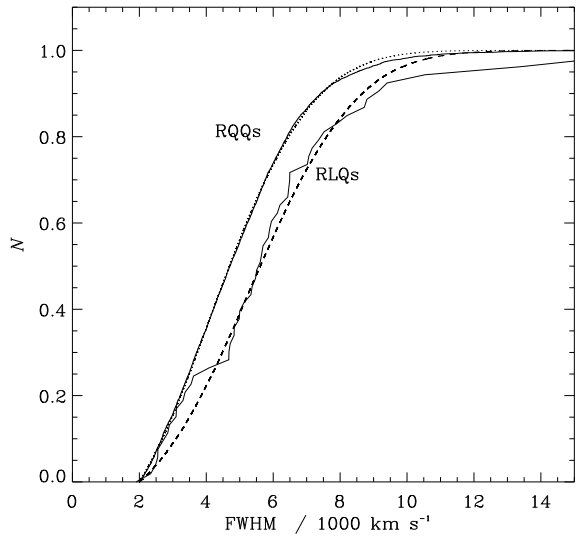}
{\caption{\label{fig:rqqrlq_cum} The best-fit models to the cumulative distribution in FWHM for the RQQs (left) and the RLQs (right). The distributions from the data are represented by the solid lines and the models are the dotted (RQQs) and dashed line (RLQs).
The best-fit model for the RLQs was obtained by freezing $\sigma_{\rm orb }$ and $\theta_{\rm max}$ at the same values found for the RQQs (see text for details).}}
\end{figure}

\subsection{Do radio-quiet quasars have orientation dependent broad lines?}\label{sec:rqq}

The unique way in which our sample is defined means that we are able
to directly relate our findings for the radio-loud quasars to the
larger radio-quiet quasar (RQQ) population which have been selected
from the same optical survey, namely the SDSS. In McLure \&
Jarvis (2004) we ensured that the radio-quiet and radio-loud quasar
samples possessed the same distribution in both redshift, bolometric
luminosity and photometric colour.

However, the mean FWHMs and standard errors of the RQQs and RLQs from our sample are $4831 \pm 25$~km~s$^{-1}$ and $5858 \pm
129$~km~s$^{-1}$ respectively. This was predominantly the reason for
the difference in the estimates of the black-hole masses for RQQs and RLQs
discussed in McLure \& Jarvis (2004), where black-hole mass $M_{\rm bh} \propto $~FWHM$^{2}$.

With the result that broad-line widths are dependent on spectral index
and therefore source orientation, then it is interesting to
investigate whether the distribution in broad-line FWHM is due to a
disc-like geometry for the broad-line region.  Wills \& Browne (1986)
quantified the broad-line region as comprised of a random isotropic
velocity component $v_{r}$ and a second which is only directed in the
place of a disc $v_{p}$.  In this scenario, the observed FWHM is given
by ${\rm FWHM} \simeq 2 (v_{r}^{2} +
v_{p}^{2}~\sin^{2}\theta)^{1/2}$,where $\theta$ is the angle between
the disc normal and the line of sight to the observer. McLure \&
Dunlop (2002) used a three-parameter model where the intrinsic
distribution of orbital velocities in the disc plane is assumed to be
Gaussian, with mean $V_{\rm orb}$ and variance $\sigma^{2}_{\rm
orb}$. The third parameter in this model is the maximum opening angle
to the line-of-sight, $\theta_{\rm max}$. The distribution in the
opening angle is completely random within the solid angle defined
between $\theta=0^{\circ} \rightarrow \theta_{\rm max}$, where
$\theta=0^{\circ}$ represent a source that is pole-on along the
line-of-sight.

In Fig.~\ref{fig:rqqrlq_cum} we plot the cumulative distribution in
FWHM for the complete sample of radio-quiet quasars from McLure \&
Jarvis (2004). We enforce a minimum FWHM cut-off of 2000~km~s$^{-1}$
to ensure that any incompleteness at these line widths are minimised
within the SDSS quasar catalogue\footnote{We note that the formal FWHM
limit for the SDSS to classify an object as a quasar is
1000~km~s$^{-1}$.}.  Using the disc model, outlined above, to
reproduce the cumulative distribution of the radio-quiet quasars drawn
from our sample we find a best fit model with $V_{\rm orb} =
4700$~km~s$^{-1}$, $\sigma_{\rm orb} = 1800$~km~s$^{-1}$ and
$\theta_{\rm max} = 38.5^{\circ}$. To test whether such a disc model
is fair representation of the data we use the 1-D KS-test which shows
that the probability of the data being drawn from the best-fit model is 40 per cent. The largest difference between
the data and the model occurs at the large FWHMs, where additional
processes may alter the distribution. The most relevant of these would
be a luminosity dependent torus opening angle, i.e. the receding
torus model (Lawrence 1991). However, as the data and model are consistent we do not include more free parameters to include a description of a receding torus model.

Under the assumption that the disc-like model is a fair representation
of the true nature of the broad-line regions in quasars, then we are
able to constrain the actual difference in orbital velocity for RQQs
and RLQs.  As we have no reason to believe that the distribution of
angles of the disk and the opening angle should be different in RQQs
and RLQs, we use the best-fit parameters of $\sigma_{\rm orb}$ and
$\theta_{\rm max}$ for the RQQs, and leave $V_{\rm orb}$ as a free
parameter to fit the whole RLQ population (both FSQs and SSQs). We
find a best-fit of $V_{\rm orb} = 5950$~km~s$^{-1}$ for the RLQs,
shown in Fig.~\ref{fig:rqqrlq_cum}.  The KS-test shows that the
probability of the data being drawn from this model is 68 per cent,
and is therefore consistent.  However, one can immediately see an
excess of objects with FWHM $< 5000$~km~s$^{-1}$ in the RLQ data when
compared to the best-fit model distribution.

It is also worth noting that the distribution in FWHM for the RQQs is
indistinguishable from that of the FSQs, with the KS-test giving a
probability of 75 per cent that they are drawn from the same
underlying distribution. On the other hand the KS-test shows that the
probability of the SSQs being drawn from the same distribution as the
RQQs is $\ll0.01$ per cent.

We defer discussion of the excess number of RLQ sources at FWHM $< 5000$~km~s$^{-1}$ to the next section, where we also discuss the reason for the
agreement between the FWHM distributions of the RQQs and FSQs and
their observed difference with the FWHMs of the SSQ broad-lines.

\section{Discussion}\label{sec:discuss}

\subsection{Flat-spectrum quasars as beamed radio-quiet quasars}\label{sec:beamed}

One possible explanation for the similarity in the FWHM distributions
for RQQs and FSQs is that the FSQ population is predominantly
comprised of RQQs.  There is evidence to suggest that radio-quiet
quasars have large-scale jets, similar to those found in FRI
(Fanaroff \& Riley 1974) radio galaxies (Blundell \& Rawlings 2001).
Due to the orientation of the source, the intrinsic low-luminosity of
the radio jet may be Doppler boosted towards the observer. This was
discussed fully in Jarvis \& McLure (2002) where
consideration of both the orientation and Doppler boosting of the radio
emission was used to explain the apparent discrepancy of FSQs
(Oshlack, Webster \& Whiting 2002) on the black-hole mass -- radio
luminosity correlation put forward by a number of authors
(Franceschini, Vercellone \& Fabian 1998; Dunlop et al. 2003; McLure
et al. 2004).

To estimate the amount of boosting that would be required to
place the RQQs in the radio-loud domain at $L_{1.4} >
10^{24}$~W~Hz$^{-1}$~sr$^{-1}$ (Miller, Peacock \& Mead 1990), we use
the average luminosity of the FSQs, $L_{1.4} = 8.7
\times 10^{25}$W~Hz$^{-1}$~sr$^{-1}$. In order to boost these sources from the
RQQ domain then Doppler factors of $\Gamma^{2} \sim 10-100$ are
required. This is a relatively small amount of boosting. For example,
as discussed in Jarvis \& McLure (2002) factors of $\sim 100-1000$ are
easily obtainable under reasonable assumptions of bulk Lorentz factor
and, in the case of powerful FSQs, a maximum opening angle of
$\theta_{\rm max} \sim 7^{\circ}$ (Jackson \& Wall 1999). For the
larger opening angle of $\theta_{\rm max} \sim 40^{\circ}$ from our
modelling in section~\ref{sec:rqq}, then this would obviously reduce the average
amount of Doppler boosting\footnote{The Doppler factor $\Gamma$ is
given by $\Gamma =
\gamma^{-1} (1 -
\beta\cos\theta)^{-1}$,with $\beta= v/c$, where $v$ the speed of the
bulk motion, $\theta$ is the angle of the jet to the line of sight,
and $\gamma$ is the bulk Lorentz factor.}.  However, a significant
fraction of sources would still exhibit substantial amounts of Doppler
boosting as they would lie within $\sim 20^{\circ}$ of our
line-of-sight, which may still result in $\Gamma^{2} \gtsim 10$ as the distribution in angles follows $1-\cos(\theta)$, i.e. proportional to the solid angle.
Therefore, given a maximum opening angle of $\theta \sim 40^{\circ}$
then, for a disc-like broad-line region, we would
expect $10 - 25$ per cent of the RLQ population to exhibit Doppler
factors of $\Gamma^{2} > 10$, i.e. those within $\theta <
20^{\circ}$.

Based on this reasoning, as many as half of the FSQs in our
sample could be Doppler boosted RQQs. This is qualitatively consistent with Fig.~\ref{fig:rqqrlq_cum} where there is an obvious ridge in the cumulative distribution of RLQs where $\sim 20$ per cent of the sample deviate from the best fit model at $\ltsim 5000$~km~s$^{-1}$.

It is worth noting that the radio-frequency at which the RLQs are
selected is the crucial factor which dictates the fraction of
flat-spectrum, Doppler boosted sources, in any RLQ
sample. For example, the sample used in Oshlack et al. (2002) was
derived from the 2.7~GHz Parkes flat-spectrum sample of Drinkwater et
al. (1997). Therefore, the Oshlack et al. sample would contain a much higher fraction of
Doppler boosted sources than the sample discussed in this paper, due to the fact that by definition it is comprised solely of flat-spectrum sources. Whereas
the need for a low-frequency measurement from the WENSS means that the
fraction of Doppler boosted sources in our sample is biased towards a lower
value. This is due to the fact that many lower luminosity,
flat-spectrum radio sources which are detected in FIRST will not be
included here as they fall below the WENSS flux-density
limit. Therefore, we are enforcing an orientation independent selection on the optical quasars in the SDSS by using the WENSS flux-density limit at 325~MHz.

\subsection{Consequences for the virial black-hole mass estimator}\label{sec:bh}

In this section we consider how the scenario outlined in
section~\ref{sec:beamed} would affect the estimation of the black-hole
masses in quasars via the virial method (e.g. Wandel, Peterson \&
Malkan 1999; Kaspi et al. 2000; McLure \& Jarvis 2002).

The argument that the FSQs comprise objects that are Doppler boosted
from the RQQ regime does fit in with orientation based unified
schemes. However, it is important to note that this argument is based
on the {\it average} properties of the FSQ population. Some of them
will no doubt be bona fide radio-loud quasars with little or no
Doppler boosting, and the fraction of such sources will be heavily
dependent on the flux-density limit and frequency of the survey from
which the sample is derived, as discussed in section~\ref{sec:beamed}.

With the disc model for the RQQ and RLQ populations determined in
section~\ref{sec:rqq}, then the difference in black-hole mass of RQQs
and RLQs is 0.18~dex (as $M_{\rm bh} \propto V_{\rm orb}^{2}$),
consistent with the 0.16~dex found by assuming FWHM$\equiv V_{\rm
orb}$ in the analysis of McLure \& Jarvis (2004), where no
consideration of source orientation was made. The sample of RLQs used
in McLure \& Jarvis comprised all of the radio sources with detections
in FIRST, without the need for a WENSS detection, thus it is highly
likely that the fraction of flat-spectrum sources is larger in the
complete sample. Unfortunately this is currently impossible to test
due to the lack of low-frequency radio data for all of the sources.
Moreover, if higher-resolution, multi-frequency radio data were
available for the sample used here then it is likely that the
difference in black-hole mass for the RQQs and genuine RLQs would
increase as the beamed RQQs could be removed from the RLQ sample.
This would then push the difference closer to the value found from
modelling of the host galaxies of radio galaxies and quasars.
At low redshift ($0.1 < z <
0.25$), Dunlop et al. (2003) showed that the black-hole masses derived
using the relation between black-hole mass and spheroid luminosity
(e.g. Magorrian et al. 1998; Gebhardt et al. 2000; Merritt \&
Ferrarese 2001) were significantly lower in radio-quiet quasars, with
a mean difference of $\sim 0.22$~dex. 
The radio luminosity dependence on black-hole mass is also consistent
with results from modelling the host galaxies of powerful radio
galaxies. Using the black-hole mass -- bulge luminosity relation to
estimate black-hole mass, McLure et al. (2004) showed that the radio
luminosity of powerful FRII-type (Fanaroff \& Riley 1974) radio
galaxies is correlated with the bulge luminosity and hence the
black-hole mass, at least at $z \sim 0.5$.

Therefore, our analysis reinforces the view that {\it intrinsic} radio power is dependent
on the black-hole mass. 
However, there will be a coupling with extra
parameters such as accretion rate and/or black-hole spin, along with
the orientation effects discussed here.

This can also be seen in the principal component analysis (PCA) of Boroson
(2002). The median value for principal component 1, which is
believed to represent the accretion rate, for FSQs is shifted closer
to the region in which the RQQs lie when compared to the SSQs. This
implies that, even in these radio-selected samples, FSQs and RQQs
may be drawn from the same population within $\sim 1$~dex of the
boundary of radio-loud and radio-quiet classification. Unfortunately,
we do not have enough sources in our sample at the correct redshift to carry
out the PCA discussed in Boroson (2002). However deeper low-frequency
radio observations of the quasars over the FIRST area would allow us
to increase the sample size at low redshift and thus carry out the
PCA.

We have shown that the FWHM of broad lines measured in both the RLQs {\it and} the RQQs will have an orientation dependency. Therefore, measuring black-hole masses via the
virial estimate on small samples may lead to spurious
results. However, the scatter due to the source orientation should be
overcome if large samples of quasars are used, and this is evidently possible with quasar surveys such as the SDSS and 2dfQSO (Croom et al. 2004).

Decoupling orientation effects and black-hole mass differences will be
a difficult task. However, deep high-resolution multi-frequency radio
observations of large samples of RQQs will provide the best way of placing
direct constraints on this problem.

\section{Summary}\label{sec:summary}

We have used our well defined sample of radio-loud and radio-quiet quasars to investigate the dependence of the observed broad-line widths on orientation. To summarise

\begin{itemize}

\item Radio-quiet quasars and the flat-spectrum quasars exhibit the same distribution in broad-line widths, whereas the steep-spectrum population exhibit a distribution which is shifted significantly to broader widths. 

\item The difference in FWHM distributions can be explained within unified schemes where a substantial fraction of the flat-spectrum sources are composed of Doppler boosted radio-quiet quasars. For the sample we use in this paper, this fraction could be as high as 50 per cent of the flat-spectrum quasars and 25 per cent of all radio-loud quasars.

\item In light of this we agree with previous results which suggest that radio-quiet quasars contain less massive black holes than their radio-loud quasar counterparts. We find a difference on the mean black-hole masses of 0.18~dex between radio-loud and radio quiet quasars. However, this is likely to be a lower limit due to the difficulty in accounting for beamed radio-quiet quasars which lie above the divide which separates radio-loud and radio-quiet quasars.

\item Under this hypothesis, the measured broad-line widths are dependent on both orientation and black-hole mass. The level of orientation bias will be heavily dependent on the selection method, thus studies using the black-hole masses derived from the virial estimate need well defined quasar samples where any possible orientation bias is kept to a minimum.

\item We suggest that a method of decoupling the dependencies of orientation and black-hole mass and to keep orientation effects to a minimum would be to carry out deeper low-frequency radio observation across the FIRST area covered by the SDSS. Such a survey could be easily achieved with future radio telescopes such as the low-frequency array (LOFAR; http://www.lofar.org).

\end{itemize}

\section*{ACKNOWLEDGEMENTS} 

MJJ acknowledges funding from a PPARC PDRA and RJM acknowledges
funding from the Royal Society. We thank the referee Katherine Inskip
for a detailed reading of the manuscript.
This publication makes use of the material provided in the FIRST, NVSS
and SDSS surveys. FIRST is funded by the National Radio Astronomy
Observatory (NRAO), and is a research facility of the US National
Science foundation and uses the NRAO Very Large Array.  Funding for
the creation and distribution of the SDSS Archive has been provided by
the Alfred P. Sloan Foundation, the Participating Institutions, the
National Aeronautics and Space Administration, the National Science
Foundation, the U.S. Department of Energy, the Japanese
Monbukagakusho, and the Max Planck Society. The SDSS Web site is
http://www.sdss.org/. The SDSS is managed by the Astrophysical
Research Consortium (ARC) for the Participating Institutions. The
Participating Institutions are The University of Chicago, Fermilab,
the Institute for Advanced Study, the Japan Participation Group, The
Johns Hopkins University, the Korean Scientist Group, Los Alamos
National Laboratory, the Max-Planck-Institute for Astronomy (MPIA),
the Max-Planck-Institute for Astrophysics (MPA), New Mexico State
University, University of Pittsburgh, University of Portsmouth,
Princeton University, the United States Naval Observatory, and the
University of Washington.

{} 

\end{document}